\documentclass[sigconf]{acmart}

\usepackage{changepage}

\DeclareMathOperator*{\minimize}{minimize}
\usepackage{xcolor}
\usepackage{bbm}
\usepackage[ruled,vlined]{algorithm2e}
\usepackage{multirow}
\usepackage{subcaption}

\AtBeginDocument{%
  \providecommand\BibTeX{{%
    \normalfont B\kern-0.5em{\scshape i\kern-0.25em b}\kern-0.8em\TeX}}}

\theoremstyle{theorem}

\theoremstyle{proof}

\copyrightyear{2021} 
\acmYear{2021} 
\setcopyright{acmlicensed}\acmConference[CIKM '21]{Proceedings of the 30th ACM International Conference on Information and Knowledge Management}{November 1--5, 2021}{Virtual Event, QLD, Australia}
\acmBooktitle{Proceedings of the 30th ACM International Conference on Information and Knowledge Management (CIKM '21), November 1--5, 2021, Virtual Event, QLD, Australia}
\acmPrice{15.00}
\acmDOI{10.1145/3459637.3482420}
\acmISBN{978-1-4503-8446-9/21/11}

\begin{document}
\fancyhead{}
\title{Counterfactual Explainable Recommendation}
\author{Juntao Tan}
\affiliation{%
  \institution{Rutgers University}
  \city{New Brunswick, NJ}
  \country{US}
}
\email{juntao.tan@rutgers.edu}

\author{Shuyuan Xu}
\affiliation{%
  \institution{Rutgers University}
  \city{New Brunswick, NJ}
  \country{US}
}
\email{shuyuan.xu@rutgers.edu}

\author{Yingqiang Ge}
\affiliation{%
  \institution{Rutgers University}
  \city{New Brunswick, NJ}
  \country{US}
}
\email{yingqiang.ge@rutgers.edu}

\author{Yunqi Li}
\affiliation{%
  \institution{Rutgers University}
  \city{New Brunswick, NJ}
  \country{US}
}
\email{yunqi.li@rutgers.edu}

\author{Xu Chen}
\affiliation{%
  \institution{Renmin University of China}
  \city{Beijing}
  \country{China}
}
\email{xu.chen@ruc.edu.cn}

\author{Yongfeng Zhang}
\affiliation{%
  \institution{Rutgers University}
  \city{New Brunswick, NJ}
  \country{US}
}
\email{yongfeng.zhang@rutgers.edu}
\begin{abstract}
By providing explanations for users and system designers to facilitate better understanding and decision making, explainable recommendation has been an important research problem. In this paper, we propose \textbf{Count}erfactual \textbf{E}xplainable \textbf{R}ecommendation (CountER), which takes the insights of counterfactual reasoning from causal inference for explainable recommendation. CountER is able to formulate the complexity and the strength of explanations, and it adopts a counterfactual learning framework to seek simple (low complexity) and effective (high strength) explanations for the model decision. Technically, for each item recommended to each user, CountER formulates a joint optimization problem to generate minimal changes on the item aspects so as to create a counterfactual item, such that the recommendation decision on the counterfactual item is reversed. These altered aspects constitute the explanation of why the original item is recommended. The counterfactual explanation helps both the users for better understanding and the system designers for better model debugging.

Another contribution of the work is the evaluation of explainable recommendation, which has been a challenging task. Fortunately, counterfactual explanations are very suitable for standard quantitative evaluation. To measure the explanation quality, we design two types of evaluation metrics, one from user's perspective (i.e. why the user likes the item), and the other from model's perspective (i.e. why the item is recommended by the model). We apply our counterfactual learning algorithm on a black-box recommender system and evaluate the generated explanations on five real-world datasets. Results show that our model generates more accurate and effective explanations than state-of-the-art explainable recommendation models. Source code is available at \url{https://github.com/chrisjtan/counter}.
\end{abstract}

\ccsdesc[500]{Computing methodologies~Machine learning}
\ccsdesc[500]{Information systems~Recommender systems}

\keywords{Explainable Recommendation; Counterfactual Explanation; Counterfactual Reasoning; Machine Learning; Explainable AI}

\maketitle

\vspace{-5pt}
\section{Introduction}

Explainability for recommender systems is crucial, because in recommendation scenarios we can rarely say that some recommendation is absolutely right or some other recommendation is absolutely wrong, instead, 
it all depends on good explanations to help users understand why an item is recommended so as to increase the transparency and trust and enable better decision making; good explanations also help system designers to track the behavior of the complicated recommendation models for better debugging \cite{Zhang2020,zhangsigir14}.

One prominent approach is aspect-aware explainable recommendation \cite{zhangsigir14, Wang2018, chen2016, chen2020try}, which takes the explicit item features/aspects to construct explanations. For example, \citeauthor{zhangsigir14} \cite{zhangsigir14} proposed Explicit Factor Model (EFM) which aligns latent factors with explicit features such as color and price to generate sentence explanations in the form of ``You might be interested in [feature], on which this product performs well.'' \citeauthor{Wang2018} \cite{Wang2018} learned the user-aspect preferences in a multi-task joint tensor factorization framework to construct the aspect-aware explanations. \citeauthor{chen2020try} \cite{chen2020try} explored attribute-aware collaborative filtering for explainable substitute recommendation. \citeauthor{li2021personalized} \cite{li2021personalized} proposed a Personalized Transformer to generate aspect-inspired natural language explanations.
A more comprehensive review of related work is provided in Section \ref{sec: related}. 


However, existing methods on aspect-aware explainable recommendation face several issues: 
1) Most of the methods are designed as intrinsic explainable models, although they have the advantage of providing faithful explanations, it may be difficult for them to explain other black-box recommendation models.
2) Existing methods do not consider the explanation complexity, in particular, they do not have the ability to decide how many aspects to use when generating explanations. Most methods generate explanations using exactly one aspect, however, the real reason of the recommendation may be triggered by multiple aspects. 3) Existing methods do not consider the explanation strength, i.e., to what extent the explanation really influences the recommendation result. This is mostly because existing methods are designed based on matching algorithms, which extracts associative signals such as feature importance and attention weights to construct explanations, while they seldom consider what happens if we intervene these signals to alternative values.

\begin{figure}
    \centering
    \includegraphics[width=1\linewidth]{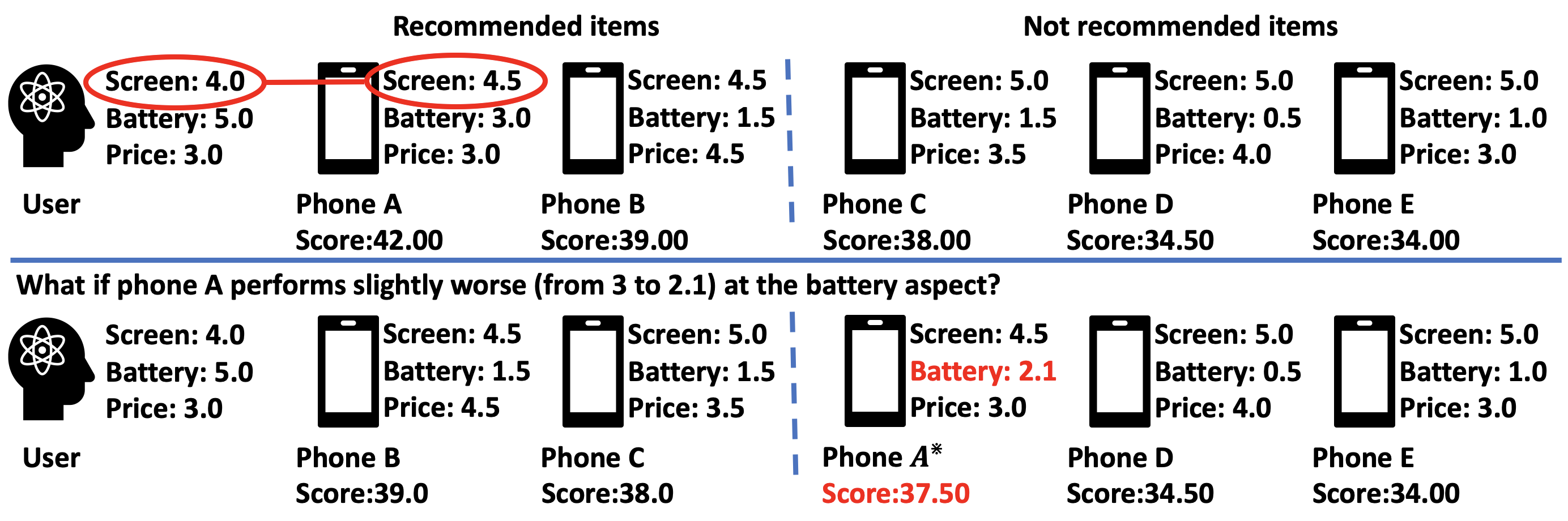}
    \vspace{-20pt}
    \caption{Matching-based vs. counterfactual reasoning. The numbers beside the three aspects (screen, battery, and price) show how much the user cares about an aspect and how well the item performs on an aspect. In this example, matching-based explanation methods would use ``screen'' to construct an explanation, while counterfactual reasoning method will use ``battery'' to construct the explanation.}
    \vspace{-10pt}
    \label{fig:overview}
\end{figure}

Recent advances on counterfactual reasoning shed light on the possibility to solve the above problems. We use a toy example in Figure \ref{fig:overview} to illustrate the intuition of counterfactual explanation and its difference from matching-based explanation.
In this example, similar to \citeauthor{zhangsigir14} \cite{zhangsigir14}, each user has his/her preference score on each aspect, while each item has its performance score on each aspect. The recommendation algorithm uses the total score to rank items. For example, the total score of Phone A is $4.0 \times 4.5 + 5.0 \times 3.0 + 3.0\times 3.0=42$, which ranks at the 1st position of the recommendation list. To explain the recommendation of Phone A, a matching-based model would select \textit{screen} as the explanation because the multiplication score for \textit{screen} ($4.0 \times 4.5 = 18$) is the highest compared to the other aspects ($5.0 \times 3.0=15$ for \textit{battery} and $3.0\times 3.0 = 9$ for \textit{price}). However, Phone A actually performs the worst on \textit{screen} among all products, which makes this explanation unreasonable. 
This shows that the aspects with high matching scores may not always be the reason of the recommendation. 

Counterfactual reasoning, on the contrary, aims to understand the underlying mechanism that really triggered the recommendation by applying interventions on the aspects and see what happens. As shown in the lower part of Figure \ref{fig:overview}, we can look for the minimal change on Phone A's features such that Phone A will not be recommended anymore. In this case, the aspect \textit{battery} will be selected as explanation because we only need to slightly change its score from 3 to 2.1 to reverse the recommendation decision, indicating that \textit{battery} is a very influential factor on the recommendation result.


Following the above intuition, in this paper, we propose Counterfactual Explainable Recommendation (CountER), which adopts counterfactual reasoning to extract aspect-aware explanations for recommendation. Inspired by the Occam's Razor Principle \cite{blumer1987occam}, CountER is built on the fundamental idea of explanation complexity and explanation strength, where complexity measures how much change must be applied on the factual observations, and strength shows to what extent the applied change will influence the recommendation decision. Through our proposed counterfactual constrained learning framework, CountER aims to extract simple (low complexity) and effective (high strength) explanations for the recommendation by looking for minimal changes on the facts that would alter the recommendation decision.

Another important challenge in explainable recommendation (and explainable AI) research is how to evaluate the explanations. Due to the lack of standard offline evaluation measures for explanation, previous research heavily relied on human subjects for evaluation, which makes the evaluation process expensive, unscalable, hard to standardize, and unfriendly to academic research settings \cite{Zhang2020}. Fortunately, counterfactual explanations are very suitable for standard quantitative evaluation. In this paper, we propose two types of evaluation methods, one is user-oriented evaluation, and the other is model-oriented evaluation. 
For user-oriented evaluation, we adopt similar ideas as \cite{le2020,li2020generate,li2021personalized} by using each user's mentioned aspects in their reviews as the ground-truth. By comparing our generated explanation with the ground-truth, we can evaluate the feature coverage, precision, recall, and $F_1$ scores.
For model-oriented evaluation, 
we adopt the Probability of Necessity (PN), Probability of Sufficiency (PS) and their harmonic mean $F_{NS}$ to evaluate the sufficiency and necessity of the explanations. More details are provided in the experiments.

In summary, this work has the following contributions:
\vspace{-1ex}
\begin{enumerate}
    \item For the first time, we explore the complexity and the strength of explanations in explainable recommendation and formulate the concepts in mathematical ways.
    \item We formulate a counterfactual reasoning framework based on counterfactual constrained learning to extract simple and effective explanations for recommendation.
    \item We design both user-oriented and model-oriented metrics for standard evaluation of explainable recommendation. 
    \item We conduct extensive experiments on five real-world datasets to validate the effectiveness of our proposed method.
\end{enumerate}



\section{Related Work}

\label{sec: related}
In this section, we review some related work on explainable recommendation and counterfactual reasoning.

\subsection{Explainable Recommendation}

Explainable recommendation is a broad research area with many different types of models, and it is difficult to cover all of the works on this direction. Since our work is more closely related with aspect-aware explainable recommendation, we mainly focus on this sub-area in the section. A more complete review of explainable recommendation can be seen in \cite{Zhang2020,IJHCS14-HCI,Handbook15-Explanation}.

Explainability of recommender systems is important because it improves the transparency, user satisfaction and trust over the recommendation system \cite{Zhang2020}. One representative way to generate user-friendly explanations is by modeling aspects in the recommended items. For instance, \citeauthor{zhangsigir14} \cite{zhangsigir14} introduced an Explicit Factor Model (EFM) for explainable recommendation.
It first extracts the item aspects and the user opinions on these aspects from user reviews. Then, it trains a matrix factorization-based recommendation model to generate aspect-aware explanations by aligning the latent factors with the item aspects. \citeauthor{chen2016} \cite{chen2016} and \citeauthor{Wang2018} \cite{Wang2018} advanced from matrix factorization to tensor factorization models for explainable recommendation. \citeauthor{he2015} \cite{he2015} proposed a tripartite graph model to improve the interactivity of aspect-aware recommendation models. \citeauthor{gao2019explainable} \cite{gao2019explainable} proposed an explainable deep model to learn multi-level user profile and infer which level of features best captures a user's interest. \citeauthor{balog2019transparent} \cite{balog2019transparent} presented a set-based recommendation technique to improve the scrutability and transparency of recommender systems. \citeauthor{ren2017social} \cite{ren2017social} proposed a collaborative viewpoint regression for explainable social recommendation. \citeauthor{wang2018tem} \cite{wang2018tem} proposed a tree-enhanced embedding method to combine embedding-based and tree-based models for explainable recommendation. More recently, \citeauthor{chen2020try} \cite{chen2020try} applied a residual feed-forward neural network to model the user and item explicit features and generates explainable substitute recommendations. \citeauthor{pan2020explainable} \cite{pan2020explainable} presented a feature mapping approach to
map the latent features onto the interpretable aspects to achieve both satisfactory accuracy and explainability. \citeauthor{li2021personalized} \cite{li2021personalized} proposed a Personalized Transformer to generate aspect-inspired natural language explanations. \citeauthor{xian2021ex3} \cite{xian2021ex3} developed an attribute-aware algorithm for explainable item-set recommendation and deployed in Amazon. Some other explainable recommendation methods include knowledge graph-based explanations \cite{xian2019reinforcement,10.1145/3340531.3412038,ai2018learning}, neural logic explanations \cite{zhu2021faithfully}, visual explanations \cite{chen2019personalized}, natural language explanations \cite{li2020generate,chen2019generate,li2021personalized,li2017neural}, dynamic explanations \cite{chen2019dynamic}, reinforcement learning-based explanations \cite{wang2018reinforcement}, conversational explanations \cite{chen2020towards,zhao2019personalized}, fair explanations \cite{fu2020fairness}, disentangled explanations \cite{ma2019learning}, review-based explanations \cite{chen2018neural,ni2019justifying}, etc.


Most of the existing approaches generate explanations based on a very similar hypothesis: If there exists an aspect that best matches between the user's preference and the item's performance, 
then this aspect would be the explanation of the recommendation.
However, our method generates explanations from a counterfactual perspective: if an item would not have been recommended had it performed slightly worse on some aspects, then these aspects would be the reason for the model to recommend this item.



\vspace{-5pt}
\subsection{Counterfactual Reasoning}
Counterfactual reasoning, together with logical reasoning \cite{shi2020neural,chen2021neural}, are two important types of cognitive reasoning approaches. Recently, counterfactual reasoning has drawn attention in explainable AI research. It has some successful applications in several machine learning fields such as computer vision \cite{goyalarxiv2019}, natural language processing \cite{hendricks2018, feder2020}, and social fairness \cite{wachterhjlt2018, Dandllncs2020}.
In the recommendation field, recent works used counterfactual reasoning to improve both recommendation accuracy \cite{wang2021counterfactual,xu2021causal} and explainability \cite{Ghazimatinwsdm2020,xu2021learning,tran2021counterfactual} based on heterogeneous information networks \cite{Ghazimatinwsdm2020}, perturbation model \cite{xu2021learning} or influence functions \cite{tran2021counterfactual}, e.g., \citeauthor{Ghazimatinwsdm2020} \cite{Ghazimatinwsdm2020} tried to generate provider-side counterfactual explanations by looking for a minimal set of user's historical actions (e.g. reviewing, purchasing, rating) such that the recommendation can be changed by removing the selected actions. \citeauthor{tran2021counterfactual} \cite{tran2021counterfactual} adopted influence functions for identifying training points most relevant to a recommendation while deducing a counterfactual set for explanations.

A common factor between our work with prior work is that both of the proposed methods generate explanations based on extracted causalities rather than associative relationships. Yet, our work is different from prior works on two key points:
1) In terms of problem definition, prior works generate counterfactual explanations on the user-side based on user actions while our method generates counterfactual explanations on the item-side based on item aspects, which are two different types of explanations. 
2) In terms of technique, our method adopts a counterfactual learning framework driven by the Occam's Razor Principle \cite{blumer1987occam} to directly learn an explanation of small complexity and large strength, so that our desire of finding simple and effective explanation is directly encoded into the model objective.

\section{Problem Formulation}
\label{sec: formulation}

In this section, we first describe the preliminaries and the counterfactual explainable recommendation problem. Then, we introduce the concepts of explanation complexity, explanation strength, and their relations. Finally, we introduce the intuition of counterfactual reasoning. We leave the more formal and mathematical definition of our counterfactual reasoning framework to the next section.


\subsection{Preliminaries and Notations}

Suppose we have a user set with $m$ users $\mathcal{U}=\{u_1, u_2, \cdots, u_m\}$ and an item set with $n$ items $\mathcal{V}=\{v_1, v_2, \cdots, v_n\}$. Let binary matrix $B \in \{0, 1\}^{m\times n}$ be the user-item interaction matrix, where $B_{i,j} = 1$ if user $u_i$ interacted with item $v_j$; otherwise, $B_{i,j} = 0$. We use $\mathcal{R}(u,K)$ to represent the top-$K$ recommendation list for a user $u$, and we say $v \in \mathcal{R}(u,K)$ if item $v$ is recommended to user $u$ in the user's top-$K$ list. 
Following the same method described in \citeauthor{zhangsigir14} \cite{zhangsigir14}, we apply the sentiment analysis toolkit\footnote{\url{https://github.com/evison/Sentires/}} built in \cite{zhangsentiment14} to extract (Aspect, Opinion, Sentiment) triplets from the textual reviews. For example, in the \textit{Cell Phone} domain, the extracted aspects would include \textit{color, price, screen, battery}, etc. Besides, suppose we have a total number of $r$ item aspects $\mathcal{A}=\{a_1, a_2, \cdots, a_r\}$. Same as \cite{zhangsentiment14}, we further compute the user-aspect preference matrix $X\in\mathbb{R}^{m\times r}$ and the item-aspect quality matrix $Y \in \mathbb{R}^{n\times r}$. $X_{i, k}$ indicates to what extent the user $u_i$ cares about the item aspect $a_k$. Similarly, $Y_{j, k}$ indicates how well the item $v_j$ performs on the aspect $a_k$. More specifically, $X$ and $Y$ are calculated as:
\begin{equation}
\begin{aligned}
\label{eq:XY}
    X_{i,k} &=
    \begin{cases}
    0,~\text{if user}~u_i~\text{did not mention aspect}~a_k\\
    1+(N-1)\Big(\frac{2}{1+\exp(-t_{i,k})}-1\Big),~\text{else}
    \end{cases}\\
    Y_{j,k} &=
    \begin{cases}
    0,~\text{if item}~v_j~\text{is not reviewed on aspect}~a_k\\
    1+\frac{N-1}{1+\exp(-t_{j,k}\cdot s_{j,k})},~\text{else}
    \end{cases}
\end{aligned}
\end{equation}
where $N$ is the rating scale in the system, which is 5-star in most cases. $t_{i,k}$ is the frequency that user $u_i$ mentioned aspect $a_k$. $t_{j,k}$ is the frequency that item $v_j$ is mentioned on aspect $a_k$, while $s_{j,k}$ is the average sentiment of these mentions. For both the $X$ and $Y$ matrices, their elements are re-scaled into the range of $(1,N)$ using the sigmoid function (see Eq.\eqref{eq:XY}) to match with the system's rating scale.
Since the matrix construction process is not the focus of this work, we only briefly describe this process and readers may refer to \cite{zhangsentiment14, zhangsigir14} for more details. The same user-aspect and item-aspect matrix construction technique is also used in \cite{Wang2018,gao2019explainable,le2021explainable}.


\subsection{\mbox{Counterfactual Explainable Recommendation}}

With the above definitions, the objective of our counterfactual reasoning problem is to search for aspect-driven counterfactual explanations for a given black-box recommendation model. 

More specifically, for a given recommendation model, if item $v_j$ is recommended to user $u_i$, i.e., $v_j\in\mathcal{R}(u_i,K)$, then we look for a slight change vector $\Delta=\{\delta_0, \delta_1, \cdots, \delta_r\}$ for the item-aspect quality vector $Y_j$, such that if $\Delta$ is applied on item $v_j$'s quality vector, i.e., $Y_j+\Delta$, then it will change the recommendation result to make item $v_j$ disappear from the recommendation list, i.e., $v_j\notin\mathcal{R}(u_i,K)$. All the values in $\Delta$ are either zero or negative continuous values since an item will only be removed from the recommendation list if it performs worse on some aspects. With the optimized vector $\Delta$, we can construct the counterfactual explanation for item $v_j$, which is composed of the aspects corresponding to the non-zero values in $\Delta$. The counterfactual explanation takes the following form,
\vspace{1em}
\begin{adjustwidth}{.5em}{1cm}
\fbox{\begin{minipage}{25em}
\texttt{If the item had been slightly worse on [aspect(s)], then it will not be recommended.}
\end{minipage}}
\end{adjustwidth}
\vspace{1em}
where the [aspect(s)] are selected by $\Delta$ as mentioned above. In the following, we will define the properties of $\Delta$ in more formal ways.


\subsection{Explanation Complexity and Strength}
To better understand the counterfactual explainable recommendation problem, we introduce two concepts to motivate explainable recommendation under the counterfactual reasoning background.

The first is Explanation Complexity (EC), which measures how complicated the explanation is. In our aspect-based explainable recommendation setting, the complexity can be defined as 1)
how many aspects are used to generate the explanation, which corresponds to the number of non-zero values in $\Delta$, i.e., $\|\Delta\|_0$, and 2) how many changes need to be applied on these aspects,
which can be represented as the sum of square of $\Delta$, 
i.e., $\|\Delta\|_2$. The final complexity takes a weighted sum of the two factors:
\begin{equation}
\label{eq:complexity}
C(\Delta)=\|\Delta\|_2 + \gamma\|\Delta\|_0
\end{equation}
where $\gamma$ is a hyper-parameter to control the trade-off between these two terms.

The second is Explanation Strength (ES), which measures how effective the explanation is. In our counterfactual explainable recommendation setting, this can be defined as to what extent applying the slight change vector $\Delta$ will influence the recommendation result of item $v_j$.
This can be further defined as the decrease of $v_j$'s ranking score in user $u_i$'s recommendation list after applying $\Delta$:
\begin{equation}
\label{eq:strength}
    S(\Delta) = s_{i,j} - s_{i,j_\Delta}
\end{equation}
where $s_{i,j}$ is the original ranking score of item $v_j$, and $s_{i,j_\Delta}$ is the ranking score of $v_j$ after $\Delta$ is applied to its quality vector, i.e., $Y_j+\Delta$.

We should note that Eq.\eqref{eq:complexity} and \eqref{eq:strength} are not the only way to define explanation complexity and strength. The definition depends on what we need in practice.
Our counterfactual reasoning framework introduced in Section \ref{sec: framework} is flexible and can easily adapt to different definitions of explanation complexity and strength.



It is also worth discussing the relationship between explanation complexity and strength. Actually, complexity and strength are two orthogonal concepts, i.e., a complex explanation is not necessarily strong, and a simple explanation is not necessarily weak. There may well exist explanations that are complex but weak, or simple and strong. According to the Occam's Razor Principle \cite{blumer1987occam}, if two explanations are equally effective, we prefer the simpler explanation than the complex one. As a result, counterfactual explainable recommendation aims to seek the simple (low complexity) and effective (high strength) explanations for recommendation.

\section{\mbox{Counterfactual Reasoning}}
\label{sec: framework}

In this section, we first briefly introduce the black-box recommendation model for which we want to generate explanations. Then, we describe the details of our counterfactual constrained learning framework for counterfactual explainable recommendation.

\subsection{Black-box Recommendation Model}
Suppose we have a black-box recommendation model $f$ that predicts the user-item ranking score $s_{i,j}$ for user $u_i$ and \mbox{item $v_j$ by:}
\begin{equation}
    s_{i,j}=f(X_i, Y_j\mid Z, \Theta)
\end{equation}
where $X_i$ and $Y_j$ are the user-aspect vector and item-aspect vector, respectively, as defined in Eq.\eqref{eq:XY}; $\Theta$ is the model parameter, and $Z$ represents all other auxiliary information of the model. Depending on the application, $Z$ could be ratings, clicks, text, images, etc., and $Z$ is optional in the recommendation model.
Basically, the recommendation model $f$ can be any model as long as it takes the user's and the item's aspect vectors as part of the input, which makes our counterfactual reasoning framework applicable to a wide scope of models.

In this work, to demonstrate the idea of counterfactual reasoning, we use a very simple deep neural network as the implementation of the recommendation model $f$, which includes one fusion layer followed by three fully connected layers. The network concatenates the user's and the item's aspect vectors as input and outputs a one-dimensional ranking score $s_{i,j}$. The final output layer is a sigmoid activation function so as to map $s_{i,j}$ into the range of $(0, 1)$. Then, we train the model with a cross-entropy loss:
\begin{align}\label{eqn: train base}
\begin{aligned}
    L&= -\sum\limits_{i,j, B_{i,j}=1}\log s_{i,j}-\sum\limits_{i,j, B_{i,j}=0}\log (1-s_{i,j})\\
    &= -\sum\limits_{i,j}B_{i,j}\log s_{i,j} + (1-B_{i,j})\log(1-s_{i,j})
\end{aligned}
\end{align}
where $B_{i,j}=1$ if user $u_i$ previously interacted with item $v_j$, otherwise $B_{i,j}=0$. In practice, since $B$ is a very sparse matrix, we sample the negative samples with ratio 1:2, i.e., for each positive instance we sample two negative instances. 
With this pre-trained recommendation model, for a target user, we are able to recommend top-$K$ items according to the predicted ranking scores.

\subsection{Counterfactual Reasoning}
\label{sec: counterfactual reasoning}

We build a counterfactual reasoning model to generate explanations for any item in the top-$K$ recommendation list provided by an existing recommendation model. The essential idea of the proposed explanation model is to discover slight change $\Delta$ on the item's aspects via solving a counterfactual optimization problem which is formulated in the following.


Suppose item $v_j$ is in the top-$K$ recommendation list for user $u_i$ ($v_j\in \mathcal{R}(u_i,K)$). 
As mentioned before, our counterfactual reasoning model aims to find simple and effective explanations for $v_j$, which can be shown as the following constrained optimization framework,
\begin{equation}
\begin{aligned}
    \minimize~&~\text{Explanation Complexity}\\
    \text{s.t.,}~&~\text{Explanation is Strong Enough}
\end{aligned}
\end{equation}

 Mathematically, according to our definition of explanation complexity and strength in Section \ref{sec: formulation}, the framework can be realized with the following specific optimization problem,
\begin{equation}
\label{eq:constrained_optimization}
\begin{aligned}
    \minimize~~ C(\Delta) &= \|\Delta\|_2 + \gamma\|\Delta\|_0\\
    \text{s.t.,}~~ S(\Delta) &= s_{i,j} - s_{i,j_\Delta} \ge \epsilon\\
\end{aligned}
\end{equation}
where $s_{i,j} = f(X_i, Y_j\mid  Z, \Theta)$, $s_{i,j_\Delta} = f(X_i, Y_j+\Delta\mid Z, \Theta)$. In the above equation, $s_{i,j}$
is the original ranking score of item $v_j$, $s_{i,j_\Delta}$ 
is the ranking score of $v_j$ when the slight change vector $\Delta$ is applied on $v_j$'s aspect vector.
The intuition of Eq.\eqref{eq:constrained_optimization} is trying to find an explanation $\Delta$ that is both \textit{simple} and \textit{effective}, where ``simple'' is reflected by the optimization objective, i.e., explanation complexity $C(\Delta)$ is minimized, while ``effective'' is reflected by the optimization constraint, i.e., the explanation
strength $S(\Delta)$ should be big enough to remove item $v_j$ from the top-$K$ list. 

To realize the second goal (i.e., effective/strong enough), we take the threshold $\epsilon$ as the margin between item $v_j$'s score and the $K+1$'s item's score in the original recommendation list, i.e.,
\begin{equation}
    \epsilon = s_{i,j} - s_{i,j_{K+1}}
\end{equation}
where $s_{i,j_{K+1}} = f(X_i, Y_{j_{K+1}}\mid  Z, \Theta)$ is the ranking score of the $K+1$'s item, and thus Eq.\eqref{eq:constrained_optimization} can be simplified as,
\begin{equation}
\label{eqn: not trainable}
    \begin{aligned}
    \minimize~~ \|\Delta\|_2 & + \gamma\|\Delta\|_0\\
    \text{s.t.,}~~ s_{i,j_\Delta} & \le s_{i,j_{K+1}}\\
\end{aligned}
\end{equation}

In this way, item $v_j$ will be ranked lower than the $K+1$'s item and thus be removed from the top-$K$ list. 



\subsection{Relaxed Optimization}
A big challenge to optimize Eq.\eqref{eqn: not trainable} is that both the objective $\|\Delta\|_2 + \gamma\|\Delta\|_0$ and the constraint $s_{i,j_\Delta}\le s_{i,j_{K+1}}$ are not differentiable. In the following, we relax the two parts to make the equation optimizable.

For the objective, since $\|\Delta\|_0$ is not convex, we relax it with $\ell_1$-norm
$\|\Delta\|_1$. This is shown to be efficient and provides good vector sparsity in \cite{l1_1, l1_2}, thus helps to minimize the explanation complexity in terms of the number of aspects in the explanation.


For the constraint $s_{i,j_\Delta} \le s_{i,j_{K+1}}$, we relax it as a hinge loss:
\begin{equation}
    L(s_{i,j_\Delta}, s_{i,j_{K+1}}) = \max (0, \alpha + s_{i, j_\Delta} - s_{i, j_{K+1}})
\end{equation}
and add it as a Lagrange term into the total objective.
Thus, the final optimization equation for generating explanation becomes:
\begin{align}
\label{eqn:optimizable}
\begin{aligned}
    \underset{\Delta}{\text{minimize}} ~~ \|\Delta\|_2 &+\gamma\|\Delta\|_1 + \lambda L(s_{i, j_{\Delta}}, s_{i, j_{K+1}})\\
    \text{where} ~~ s_{i,j_\Delta} = f(X_i, Y_j &+ \Delta\mid Z, \Theta),~~
    s_{i,j_{K+1}} = f(X_i, Y_{j_{K+1}}\mid  Z, \Theta)\\
    L(s_{i, j_{\Delta}}, s_{i,j_{K+1}}) &= \max (0, \alpha + s_{i, j_{\Delta}} - s_{i, j_{K+1}})
\end{aligned}
\end{align}

In Eq.\eqref{eqn:optimizable}, $\lambda$ and $\alpha$ are hyper-parameters to control the explanation strength.
A sacrifice of using relaxed optimization is that we lose the guarantee that item $v_j$ is removed from the top-$K$ list, though the probability of removing is high due to minimizing the $L(s_{i, j_{\Delta}}, s_{i, j_{K+1}})$ term. As a result, it requires a post-process to check if $s_{i,j_{\Delta}}$ is indeed smaller than $s_{i,j_{K+1}}$. We should only generate counterfactual explanations when the removal is successful. In the experiments, we will report the fidelity of our explanation model to show what percentage of items can be explained by our method. 

Besides, there is a trade-off in the relaxed model: by increasing the hyper-parameter $\lambda$, the model will focus more on the explanation strength but less on the explanation complexity.
We will also explore the influence of $\lambda$ and the relationship between explanation complexity and strength in the ablation study of the experiments.

\vspace{-5pt}
\subsection{Discussions}

\subsubsection{\textbf{Explanation Complexity for Items at Different Positions}}
\label{sec:discussion on complexity}

With the above framework, we can see that the difficulty of removing different items in the top-$K$ list are different. Suppose for a certain user, the recommender system generates top-$K$ recommended items as $v_{j_1}, v_{j_2}, \cdots, v_{j_K}$ according to the ranking scores. 
Intuitively, removing $v_{j_1}$ from the list is more difficult than removing $v_{j_K}$ from the list. The reason is that to remove $v_{j_1}$, the explanation strength should be at least $\epsilon=s_{i,j_1}-s_{i,j_{K+1}}$, which is bigger than the strength needed for removing $v_{j_K}$, which is $\epsilon=s_{i,j_K}-s_{i,j_{K+1}}$.
As a result, the generated explanations for the items at a higher position in the list will likely have higher explanation complexity, because the reasoning model has to apply larger changes or more aspects to generate high-strength explanations. 


This is a reasonable and desirable property of the counterfactual explainable recommendation framework---if the system ranks an item at a very high position, then it means that the system strongly recommends this item, which needs to be backed by strong explanation that contains more aspects. On the contrary, for an item ranked at lower positions in the list, it could be easily removed by changing only one or two aspects, which is in line with our intuition.
In the experiments, we will show the average explanation complexity for items at different positions to verify the above discussion.

\vspace{-1ex}
\subsubsection{\textbf{Controlling the Number of Aspects}}
Through Eq.\eqref{eqn:optimizable}, the model can automatically decide the number of aspects to construct the explanation. We believe this is better than choosing only one aspect as was done in previous aspect-aware explainable recommender systems \cite{zhangsentiment14, he2015, Wang2018, chen2020try}. However, if needed, our method can also generate explanations with a single aspect. To generate single aspect explanation, we adjust Eq.\eqref{eqn:optimizable} by adding a trainable one-hot vector $\textbf{a}$ as a mask to make sure that only one aspect is changed during the training. The optimization problem is:
\begin{align}
\label{eqn:single_feature}
\begin{aligned}
    \underset{\Delta,\mathbf{a}}{\text{minimize}} &~~ \|\mathbf{a}\odot \Delta\|_2 + \lambda L(s_{i, j_{\Delta}}, s_{i, j_{K+1}})\\
    \text{s.t.} &~~ \|\textbf{a}\|_1=1, a_i\in \{0, 1\}\\
    \text{where,} &~~ s_{i,j_\Delta} = f(X_i, Y_j + \mathbf{a}\odot\Delta\mid Z, \Theta)
\end{aligned}
\end{align}
Since we force the model to generate single aspect explanation, the $\ell_0$-norm term of $C(\Delta)$ vanishes because $\|\Delta\|_0=1$. 
We will explore both single- and multi-aspect explanations in experiment.

\section{Evaluating Explanations}
\label{sec:evaluation}
How to quantitatively evaluate explanations is a very important problem. Fortunately, compared to other explanation forms, counterfactual explanation is very suitable for quantitative offline evaluation. In this section, we mathematically define two types of evaluation metrics---user-oriented evaluation and model-oriented evaluation, which
we believe can help the field to move forward with standard evaluation of 
explainable recommendations.

\subsection{User-oriented Evaluation}
In user-oriented evaluation, we adopt the user's review on the item as the ground-truth reason about why the user purchased the item, which is similar to previous works \cite{li2020generate,Wang2018,li2017neural,chen2019co,chen2018neural}. 
More specifically, from the textual review that a user $u_i$ provided on an item $v_j$, we extract all the aspects that $u_i$ mentioned with positive sentiment,
which is defined as 
$P_{i,j}=\big[p_{i,j}^{(0)}, p_{i,j}^{(1)}, \cdots, p_{i,j}^{(r)}\big]$. 
$P_{i,j}$ is a binary vector, where $p_{i,j}^{(k)}=1$ if user $u_i$ has positive sentiment on the aspect $a_k$ in his/her review for item $v_j$. Otherwise, $p_{i,j}^{(k)}=0$. On the other hand, our model will produce the vector $\Delta=\{\delta_0, \delta_1, \cdots, \delta_r\}$, and those aspect(s) corresponding to the non-zero values in $\Delta$ will constitute the explanation. 

Then, for each user-item pair, we calculate the precision and recall of the generated explanation $\Delta$ with regard to the ground-truth vector $P_{i,j}$:
\begin{equation}
    \text{Precision}=\frac{\sum\nolimits_{k=1}^r p_{i,j}^{(k)}\cdot I(\delta_k)}{\sum\nolimits_{k=1}^r I(\delta_k)},~~ \text{Recall}=\frac{\sum\nolimits_{k=1}^r p_{i,j}^{(k)}\cdot I(\delta_k)}{\sum\nolimits_{k=1}^r p_{i,j}^{(k)}}
\end{equation}
where $I(\delta)$ is an identity function such that $I(\delta)=1$ when $\delta\neq0$, and $I(\delta)=0$ when $\delta=0$. In our case, the Precision measures the percentage of aspects in our generated explanation that are really liked by the user, while Recall measures how many percentage of aspects liked by the user are really included in our explanation. We also calculate the $F_1$ score as the harmonic mean between the two, i.e., $F_1=2\cdot \frac{\text{Precision}\cdot \text{Recall}}{\text{Precision} + \text{Recall}}$. Then, we average the scores of all pairs as the final Precision, Recall, and $F_1$ measure.

\subsection{Model-oriented Evaluation} 

The user-oriented evaluation only answers the question of whether the generated explanations are consistent with user's preferences. However, it does not tell us whether the explanation properly
justifies the model's behaviour, i.e., why the recommendation model recommends this item to the user.

To test if our explanation model correctly explains the essential mechanism of the recommendation system, we use two scores, Probability of Necessity (PN) and Probability of Sufficiency (PS) \cite[p.112]{pearl2016causal}, to validate the explanations with model-oriented evaluation.

In logic and mathematics, necessity and sufficiency are terms used to describe a conditional or implicational relationship between two statements. Suppose we have $S\Rightarrow N$, i.e., if $S$ happens then $N$ will happen, then we say $S$ is a sufficient condition for $N$. Meanwhile, we have the logically equivalent contrapositive $\neg N\Rightarrow \neg S$, i.e., if $N$ does not happen, then $S$ will not happen, as a result, we say $N$ is a necessary condition for $S$.

\textbf{Probability of Necessity (PN)} \cite{pearl2016causal}: In causal inference theory, probability of necessity evaluates the extent that a condition is necessary.
To calculate PN for the generated explanation, 
suppose a set of aspects $\mathcal{A}_{ij} \subset \mathcal{A}$ compose the explanation for the recommendation of item $v_j$ to user $u_i$. The essential idea of the PN score is: if in a counterfactual world, the aspects in $\mathcal{A}_{ij}$ did \textbf{not} exist in the system, then what is the probability that item $v_j$ would \textbf{not} be recommended for user $u_i$.

Following this idea, we calculate the frequency of the generated explanations that meet the PN definition. 
Let $R_{i,K}$ be user $u_i$'s original recommendation list. Let $v_j\in R_{i,K}$ be a recommended item that our framework generated a nonempty explanation $\mathcal{A}_{ij}\neq \emptyset$. Then for all the items in the universal item set $\mathcal{V}$, we alter the aspect values in the item-aspect quality matrix $Y$ to 0 if they are in $\mathcal{A}_{ij}$. In this way, we create a counterfactual item set $\mathcal{V}^*$ which results in a counterfactual recommendation list $R_{i,K}^*$ for user $u_i$ by the recommendation algorithm. Then, the PN score is:
\begin{equation}\label{eqn: PN}
    \text{PN} = \frac{\sum\nolimits_{u_i \in \mathcal{U}}\sum\nolimits_{v_j \in R_{i,K}}\text{PN}_{ij}}{\sum\nolimits_{u_i \in \mathcal{U}}\sum\nolimits_{v_j \in R_{i,K}} I(\mathcal{A}_{ij}\neq \emptyset)},
    ~\text{where}~\text{PN}_{ij}=
    \begin{cases}
      1,~\text{if $v_j^* \notin R_{i, K}^*$}\  \\
      0,~\text{else}
    \end{cases}
\end{equation}
where $I(\mathcal{A}_{ij}\neq \emptyset)$ is an identity function such that $I(\mathcal{A}_{ij}\neq \emptyset)=1$ if the condition $\mathcal{A}_{ij}\neq \emptyset$ holds and 0 otherwise. Basically, the denominator is the total number of items that the algorithm successfully generated an explanation for, and the numerator is the number of explanations that if we remove the related aspects then it will cause the item to be removed from the recommendation list. 

\textbf{Probability of Sufficiency (PS)} \cite{pearl2016causal}: 
The PS score evaluates the extent that a condition is sufficient.
The essential idea of the PS score is: if in a counterfactual world, the aspects in $\mathcal{A}_{ij}$ were the \textbf{only} aspects existed in the system, then what is the probability that item $v_j$ would still be recommended for user $u_i$.


Similarly, we calculate the frequency of the generated explanations that meet the PS definition. For all the items in $\mathcal{V}$, we alter the aspect values in the item-aspect quality matrix $Y$ to 0 if they are not in $\mathcal{A}_{ij}$.
In this way, we create a counterfactual item set $\mathcal{V}^\prime$ which results in a counterfactual recommendation list $R_{i, K}^\prime$ for user $u_i$ by the recommendation algorithm. Then, the PS score is:
\begin{equation}\label{eqn: PS}
    \text{PS} = \frac{\sum\nolimits_{u_i \in \mathcal{U}}\sum\nolimits_{v_j \in R_{i,K}}\text{PS}_{ij}}{\sum\nolimits_{u_i \in \mathcal{U}}\sum\nolimits_{v_j \in R_{i,K}} I(\mathcal{A}_{ij}\neq\emptyset)}, 
    ~\text{where}~\text{PS}_{ij}=
    \begin{cases}
      1,~\text{if $v_j^\prime \in R_{i, K}^\prime$}\  \\
      0,~\text{else}
    \end{cases}
\end{equation}
where $I(\mathcal{A}_{ij}\neq \emptyset)$ is still the identity function as above. Basically, the denominator is still the total number of items that the algorithm successfully generated an explanation for, and the numerator is the number of explanations that alone can still recommend the item to the recommendation list. Similar to the user-oriented evaluation, we also calculate the harmonic mean of PS and PN to measure the overall performance, which is $F_{NS} = \frac{2\cdot\text{PN}\cdot \text{PS}}{\text{PN}+\text{PS}}$.

\section{Experiments}
\label{sec: experiment}

In this section,
we first introduce the datasets, the comparison baselines and the implementation details. Then we present
studies on the two main expected properties in this paper: complexity and strength of the explanations. We also present ablation studies to explore how our model performs under different conditions.

\subsection{Dataset Description}
We test our method on Yelp\footnote{\url{https://www.yelp.com/dataset}} and Amazon\footnote{\url{https://nijianmo.github.io/amazon}} datasets. The Yelp dataset contains users' reviews on various kinds of businesses such as restaurants, dentists, salons, etc. The Amazon dataset \cite{ni2019justifying} contains user reviews on products in Amazon e-commerce system.
The Amazon dataset contains 29 sub-datasets corresponding to 29 product categories. We adopt four datasets of different scales to evaluate our method, which are \textit{Electronic}, \textit{Cell Phones and Accessories}, \textit{Kindle Store} and \textit{CDs and Vinyl}. 
Since the Yelp and Amazon datasets are very sparse, similar as previous work \cite{zhangsigir14,Wang2018,xian2019reinforcement,10.1145/3340531.3412038}, we remove the users with less than 20 reviews for Yelp dataset, and 10 reviews for Amazon dataset. 
The statistics of the datasets are shown in Table \ref{tab: datasets}.

\subsection{Comparable Baselines}
\label{sec:baselines}
We compare our model with three aspect-aware explainable recommendation models. We also include a random explanation baseline to show the overall difficulty of the evaluation tasks.

\textbf{EFM} \cite{zhangsigir14}: The Explicit Factor Model. This work combines matrix factorization with sentiment analysis technique to align latent factors with explicit aspects. In this way, it predicts the user-aspect preference scores and item-aspect quality scores. The top-1 aligned aspect is used to construct aspect-based explanation.

\textbf{MTER} \cite{Wang2018}: The Multi-Task Explainable Recommendation model. This work predicts a tensor $X\in \mathbb{R}^{m\times n\times (r+1)}$, which represents the affinity score among the users, items, aspects, and an extra dimension for the overall rating. This tensor $X$ is acquired via Tucker decomposition \cite{karatzoglou2010multiverse,kolda2009tensor}.
We should note that since the overall rating for a user on an item is predicted in the extra dimension via decomposition, which is not directly predicted by the explicit aspects, this method is not suitable for the model-oriented evaluation. As a result, we only report this model's explanation performance on user-oriented evaluation.

\textbf{A2CF} \cite{chen2020try}: The Attribute-Aware Collaborative Filtering model. This work leverages a residual feed-forward network to predict the missing values in the user-aspect matrix $X$ and the item-aspect matrix $Y$. The method originally considers both the user-item preference and the item-item similarity to generate explainable substitute recommendations. We remove the item-item factor to make it compatible with our problem setting to generate explanations for any item.
Similar to \cite{zhangsigir14}, the top-1 aligned aspect will be used for explanation.


\textbf{Random}: For each item recommended to a user, we randomly choose one or multiple aspects from the aspect space and generate the explanation based on them. The evaluation scores of the random baseline can indicate the difficulty of the task.

\subsection{Implementation Details}
\subsubsection{\textbf{Preprocessing}}
The preprocessing includes two parts: 1) Generating the user-aspect vector $X$ and the item-aspect vector $Y$ from the user reviews. 2) Training the base recommender system. In the preprocessing phase, we hold-out the last 5 interacted items for each user, which serve as the test data to evaluate both the recommendation and the explanation. The deep neural network in the base recommendation model consists of $1$ fusion layer and $3$ fully connected layers with \{$512$, $256$, $1$\} output dimensions, respectively. We apply ReLU activation function after all the layers except the last one, which is followed by a Sigmoid function to re-scale the predicted scores to the range of $(0,1)$. The model parameters are optimized by stochastic gradient descent (SGD) optimizer with a learning rate of $0.01$. After the recommendation model is trained, all the parameters will be fixed in the counterfactual reasoning phase and explanation evaluation phase.

\begin{table}[t]
\caption{Statistics of the datasets}
\vspace{-5pt}
\footnotesize
\setlength{\tabcolsep}{3pt}
\begin{tabular}{|l|r|r|r|r|r|}
\hline
Dataset       & \#User & \#Item & \#Review & \#Aspect & Density \\ \hline
Yelp          & 11,863 & 20,181 & 497,252       & 106        & 0.208\%         \\
CDs and Vinyl & 8,119       & 52,193       & 245,391           &  230          & 0.058\%            \\
Kindle Store  & 5,907       & 41,402       & 136,039       & 77        &    0.056\%           \\
Electronic    & 2,832       & 19,816    &    53,295           & 105           & 0.095\%              \\ 
Cell Phones   & 251       & 1,918       & 4,454           & 88           &  0.935\%             \\ 
 \hline
\end{tabular}
\vspace{-5pt}
\label{tab: datasets}
\end{table}

\begin{table}[t]
\caption{Explanation Fidelity}
\label{tab: successful rate}
\vspace{-5pt}
\footnotesize
\begin{tabular}{|l|r|r|r|r|}
\hline
\multirow{2}{*}{Datasets} & \multicolumn{2}{l|}{Single Aspect} & \multicolumn{2}{l|}{Multiple Aspect} \\ \cline{2-5} 
                          & Mask          & No Mask            & Mask           & No Mask             \\ \hline
Yelp                      & 61.60\%          & 62.54\%         & 79.30\%           & 86.38\%          \\ 
CDs and Vinyl             & 41.52\%          & 75.84\%         & 93.80\%           & 94.20\%          \\ 
Kindle Store              & 93.72\%          & 94.76\%         & 100.00\%          & 100.00\%         \\ 
Electronic                & 63.32\%          & 70.52\%         & 81.80\%           & 98.00\%          \\ 
Cell Phones               & 69.33\%          & 74.26\%         & 85.20\%           & 99.92\%          \\ \hline
\end{tabular}
\vspace{-10pt}
\end{table}


\begin{table*}[t]
\caption{User-oriented evaluation of the explanations. All numbers in the table are percentage numbers with `\%' omitted.}
\vspace{-5pt}
\label{tab:user-oriented}\small
\setlength{\tabcolsep}{4.7pt}
\begin{tabular}{|c|c|c|c|c|c|c|c|c|c|c|c|c|c|c|c|}
\hline
\multirow{3}{*}{} & \multicolumn{15}{c|}{Single Aspect Explanation}                                                                                                                                    \\ \cline{2-16} 
                  & \multicolumn{3}{c|}{Electronic} & \multicolumn{3}{c|}{Cell Phones} & \multicolumn{3}{c|}{Kindle Store} & \multicolumn{3}{c|}{CDs and Vinyl}& \multicolumn{3}{c|}{Yelp}  \\ \cline{2-16} 
                  & Pr\%      & Re\%     & F1\%     & Pr\%        & Re\%       & F1\%       & Pr\%        & Re\%        & F1\%       & Pr\%        & Re\%        & F1\%        & Pr\%         & Re\%        & F1\%        \\ \hline
Random                & 0.64          & 0.48      & 0.53         & 1.27 & 0.53    &    0.74      & 1.76          & 1.33          & 1.45          & 0.00           & 0.00          & 0.00   & 1.09 & 1.09  & 1.09        \\ \hline
EFM\cite{zhangsigir14}               & 24.58        & 16.52       & 18.79 & 20.01          & 15.33         & 17.98       & 39.25         & 34.37          & 35.96        & 36.36         & 20.00        & 24.24        & 7.18          & 7.18      & 7.18        \\ \hline
MTER\cite{Wang2018}              & 23.90       & 17.45    & 19.32       & 20.83          & 11.35         & 14.06         & 31.23          & 27.31          & 28.54         & 15.52          & 13.79          & 14.37          & 6.17           & 6.17          & 6.17          \\ \hline
A2CF\cite{chen2020try}              & 32.91      & 23.35      & 26.88      & \textbf{34.29}         & 17.71      & 22.38       & 39.38         & 35.30       & 36.59    & 33.33         & 21.67         & 24.81          & 6.33          & 6.33         & 6.33         \\ \hline \hline
CountER             & 31.58   & 23.22 & 25.60     & 25.00       & 17.40       & 19.62     & 40.45   & 35.71    & 37.24        & 25.86 & 17.30          & 19.83   & 12.73        & 12.73       & 12.73      \\ \hline
CountER  (w/ mask)            & \textbf{33.94}    & \textbf{25.67} & \textbf{28.31}     & 29.21        & \textbf{20.26}        & \textbf{22.85}      & \textbf{40.94}    & \textbf{36.19}    & \textbf{37.73}          & \textbf{39.06}     & \textbf{25.93}          & \textbf{29.33}  & \textbf{12.96}        & \textbf{12.96}       & \textbf{12.96}       \\ \hline
\multirow{3}{*}{} & \multicolumn{15}{c|}{Multiple Aspect Explanation}                                                                                                                                 \\ \cline{2-16} 
                   & \multicolumn{3}{c|}{Electronic} & \multicolumn{3}{c|}{Cell Phones} & \multicolumn{3}{c|}{Kindle Store} & \multicolumn{3}{c|}{CDs and Vinyl} & \multicolumn{3}{c|}{Yelp}\\ \cline{2-16} 
                  & Pr\%      & Re\%     & F1\%     & Pr\%        & Re\%       & F1\%       & Pr\%        & Re\%        & F1\%       & Pr\%        & Re\%        & F1\%        & Pr\%         & Re\%        & F1\%       \\ \hline
Random                    &  0.19         & 0.75      & 0.30         & 1.43   & 2.87  & 1.46         & 1.79        & 1.75       & 1.74       & 0.39           & 0.68          & 0.49  & 1.06       & 3.54       & 1.62       \\ \hline
EFM\cite{zhangsigir14}               & 19.80      &\textbf{56.56}  & 27.48    & 12.53        & 30.48        & 17.81         & 30.47         & 85.81        & 43.39        & 17.39          & 32.94      & 20.50         & 5.32          & 9.76         & 6.70     \\ \hline
MTER\cite{Wang2018}              & 10.54        & 27.24       & 13.42       & 12.50          & 25.00         & 16.67         & 12.93          & 40.16          & 18.90         & 14.41          & 39.28          & 19.74          & 5.32           & 15.93          & 7.56          \\ \hline
A2CF\cite{chen2020try}              & 18.07      & 53.72      & 25.32       & 16.05         & 28.29      & 18.76      & 31.86        & 80.44         & 43.27          & 19.39          & 57.84         & 26.62         & 4.58         & 16.67       & 7.00      \\ \hline \hline
CountER                 & 17.42  & 45.72        & 22.13      & 19.13         & \textbf{47.70}     & 25.05   & 27.79          & \textbf{86.07}  & 40.68      & 16.21          & 33.55      & 19.96    & 6.67       & 23.73       & 10.18   \\ \hline
CountER  (w/ mask)                & \textbf{25.68}  & 45.78         & \textbf{29.73}      & \textbf{21.72}          & 42.82     & \textbf{26.97}   & \textbf{32.24}          & 84.20   & \textbf{44.57}       & \textbf{20.95}          & \textbf{68.98}      & \textbf{30.00}    & \textbf{9.09}        & \textbf{28.57}       & \textbf{13.57}    \\ \hline
\end{tabular}
\vspace{-10pt}
\end{table*}

\subsubsection{\textbf{Generating Explanations}}
\label{sec: explanation}
The base recommendation model generates the top-$K$ recommendation list for each user. $K$ is set to $5$ in the experiment. We then apply the counterfactual reasoning method to generate explanations for the items in the list. The hyper-parameter $\lambda$ is set to 100 for all the datasets. The ablation study on the influence of $\lambda$ can be seen in Section \ref{sec:lambda}. We notice that the $\ell_2$-norm $\|\Delta\|_2$ and the $\ell_1$-norm $\|\Delta\|_1$ is almost in the same scale, so that we always set $\gamma$ to 1 in our model. For the margin value $\alpha$ in the hinge loss, we tested different values for $\alpha$ in $\{0.1, 0.2, 0.3,\cdots,1.0\}$ and find that the performance does not change too much for $\alpha=0.1,0.2,\cdots0.5$ and then significantly drops for $\alpha>0.5$. As a result, we set $\alpha=0.2$ throughout the experiments.
To compare with the baselines, for each recommended item, we generate both multi-aspect and single-aspect explanation through Eq.\eqref{eqn:optimizable} and Eq.\eqref{eqn:single_feature}, respectively.

\subsubsection{\textbf{Aspect Masking}}
When generating explanations, our model directly chooses aspects from the entire aspect space, which is reflected by the change vector $\Delta$. However, in the user-oriented evaluation, a strong bias exists in the user's ground-truth review, which is that a user is more possible to mention the aspects that they have mentioned before. This may result from the personal linguistic preferences. As a result, all the baseline models (EFM, MTER, A2CF) only choose aspects from the user's previously mentioned aspects to construct the explanation. To fairly compare with them in the user-oriented evaluation, we provide an adjusted version of our model. Let $M_i\in \{0, 1\}^r$ be the user $u_i$'s mask vector. $M_i$ is a binary vector, where $M_{i,k}=1$ if aspect $a_k$ is an aspect that $u_i$ cares about, i.e., $X_{i,k}\neq0$. Otherwise, $M_{i,k}=0$. We apply this mask on $\Delta$ to generate explanation by choosing aspects only from the user's preference space. which is:
\begin{align}
\label{eqn:with mask}
\begin{aligned}
    \underset{\Delta}{\text{minimize}} &~~ \|\Delta^\prime\|_2 +\gamma\|\Delta^\prime\|_1 + \lambda L(s_{i, j_{\Delta^\prime}}, s_{i, j_{K+1}})\\
    \text{where}~~ \Delta^\prime &= M_i \odot \Delta;~~s_{i, j_{\Delta^\prime}} = f(X_i, Y_j + \Delta^\prime | Z, \Theta)\\
    L(s_{i, j_{\Delta^\prime}}& , s_{i,j_{K+1}}) = \max (0, \alpha + s_{i, j_{\Delta^\prime}} - s_{i, j_{K+1}})
\end{aligned}
\end{align}

In this case, the $\Delta^\prime$ is used to generate explanation. This aspect mask can also be applied on the single-aspect formula, i.e., $\Delta^\prime = M_i \odot \mathbf{a} \odot \Delta$. 
Notice that applying the mask does not introduce the data leakage problem because the mask is calculated based on the training set.
In the evaluation, we evaluate both the original CountER model and the masked CountER model in both user-oriented and model-oriented evaluations.

\subsubsection{\textbf{Compatible with the Baselines}}
One issue in comparison with baselines is that our model is able to generate both multi-aspect and single-aspect explanations. When generating multi-aspect explanations, the model automatically decides the best number of aspects in model learning. However, the baseline methods can only generate single-aspect explanations using the top-1 aligned aspect since they do not have the ability to decide the number of aspects. Thus, to make the baseline models also comparable in multi-aspect explanation, we use our model as a guideline to tell the baseline models how many aspects should they use to generate multi-aspect explanations. For this reason, we only compare the explanations generated for the intersection of the items which are recommended by both our model and the baseline models in multi-aspect setting.

\subsubsection{\textbf{Evaluable Explanations}}
Even if the explanation model can generate explanations for all the recommended items, not all of the explanations can be evaluated from the user's perspective. This is because the user-oriented evaluation requires the user reviews as the ground-truth data. Thus, we can only evaluate the explanations generated for the ``correctly recommended'' items which appear in the test data. 
For model-oriented evaluation, all the explanations can be evaluated.

\begin{table*}[t]
\caption{Model-oriented evaluation of the explanations. All numbers in the table are percentage numbers with `\%' omitted. MTER is not suitable for the model-oriented evaluation and the reason can be found in Section \ref{sec:baselines}.}
\vspace{-5pt}
\label{tab:model-oriented}
\small
\setlength{\tabcolsep}{4.5pt}
\begin{tabular}{|c|c|c|c|c|c|c|c|c|c|c|c|c|c|c|c|}
\hline
\multirow{3}{*}{} & \multicolumn{15}{c|}{Single Aspect Explanation}                                                                                                                                    \\ \cline{2-16} 
                  & \multicolumn{3}{c|}{Electronic} & \multicolumn{3}{c|}{Cell Phones} & \multicolumn{3}{c|}{Kindle Store} & \multicolumn{3}{c|}{CDs and Vinyl} & \multicolumn{3}{c|}{Yelp} \\ \cline{2-16} 
                  & PN\%      & PS\%      &  $F_{NS}$\%     & PN\%        & PS\%        &$F_{NS}$\%        & PN\%         & PS\%        &$F_{NS}$\%         & PN\%         & PS\%         &$F_{NS}$\%         & PN\%         & PS\%         &$F_{NS}$\%         \\ \hline
Random                     & 2.05         & 2.10      & 2.07        & 3.39  & 3.50          & 3.44         & 3.16          & 2.75       & 2.94          &  1.58          & 2.03       & 1.78   & 7.52 & 10.68       & 8.82     \\ \hline
EFM\cite{zhangsigir14}               & 8.41        & 41.13       & 13.96      & 32.31         & \textbf{82.09}         & 46.37       & 6.01          & \textbf{73.84}          & 11.12        & 10.15          & 42.63          & 16.39    & 5.87           & 61.06           & 10.71        \\ \hline
A2CF\cite{chen2020try}              & 41.45        & \textbf{77.60}        & 54.03    & 36.82          & 78.68          & 50.17     & 25.66           & 65.53          & 36.88     & 25.41           & \textbf{84.51}           & 39.07        & 17.59       & \textbf{96.92}        & 29.78        \\ \hline \hline
CountER            & \textbf{65.54}  & 68.28  & \textbf{66.83}    & \textbf{74.03} & 63.30  & \textbf{68.25}      & 34.37 & 41.50  & 37.60  & 49.62        & 54.72        & 52.04  & \textbf{65.26}       & 53.25   & \textbf{58.64}       \\ \hline
CountER (w/ mask)            & 
56.73 & 62.03 & 59.26      & 70.11  & 54.71  & 61.46        & \textbf{35.39} & 46.91 & \textbf{40.34}     & \textbf{75.17}           & 49.18          & \textbf{59.46}  & 58.52       & 52.56   & 55.38       \\ \hline
\multirow{3}{*}{} & \multicolumn{15}{c|}{Multiple Aspect Explanation}                                                                                                                                 \\ \cline{2-16} 
                  & \multicolumn{3}{c|}{Electronic} & \multicolumn{3}{c|}{Cell Phones} & \multicolumn{3}{c|}{Kindle Store} & \multicolumn{3}{c|}{CDs and Vinyl} & \multicolumn{3}{c|}{Yelp} \\ \cline{2-16} 
                  & PN\%      & PS\%      & $F_{NS}$\%      & PN\%        & PS\%        & $F_{NS}$\%        & PN\%         & PS\%        &$F_{NS}$\%         & PN\%         & PS\%         &$F_{NS}$\%         & PN\%         & PS\%         & $F_{NS}$\%         \\ \hline
Random                  & 2.24          & 4.90      & 3.08         & 6.25  & 10.13          & 7.73         & 5.80          & 7.80          & 6.65          & 3.22           & 7.65          & 4.53  & 13.84        & 12.92      & 13.36      \\ \hline
EFM\cite{zhangsigir14}               & 29.65        & 84.67        & 43.92   &    52.66       & 87.98          & 65.88      & 51.72          & \textbf{96.42}          & 67.33        & 47.65        & 87.35          & 61.66     &  16.76          & 81.68           & 27.81         \\ \hline
A2CF\cite{chen2020try}              & 59.47        & 81.66        & 68.82    & 56.45         & 80.97          & 66.52  & 52.48           & 87.59       & 65.64     & 49.12     & \textbf{91.52}         & 63.93  & 41.38        & \textbf{98.28}      & 58.24    \\ \hline \hline
CountER              & \textbf{97.08}  & \textbf{96.24} & \textbf{96.66}        & \textbf{99.52}           & \textbf{98.48} & \textbf{99.00}         & \textbf{64.00}  & 79.20           &  \textbf{70.79}     & \textbf{80.89}         & 88.60          & \textbf{84.57}  & \textbf{99.91}      & 94.12 & \textbf{96.93} \\ \hline
CountER  (w/ mask)              & 77.96  & 89.26 & 83.23        & 86.62           & 91.78 & 89.13        & 60.70 & 80.10           & 69.06      & 72.47         & 67.72           & 70.01  & 96.73       & 94.39 & 95.55  \\ \hline
\end{tabular}
\end{table*}

\vspace{-6pt}
\subsection{Experimental Results}
\label{sec: results}
We first report the fidelity of our explanation method in Table \ref{tab: successful rate}, which shows for what percentage of recommended items that our method can successfully generate an explanation. We notice that the fidelity of the single-aspect version is lower than that of the multi-aspect version. This indicates that for our model, using only one aspect as the explanation may not lead to enough explanation strength to effectively explain the recommendations. However, if we allow the model to use multiple aspects, we can find strong enough explanations in most cases ($80\%\sim100\%$). The overall average number of aspects in multi-aspect explanation is $2.79$. 


We then evaluate the explanations generated by CountER (original version and masked version) and the baselines. The user-oriented evaluation is reported in Table \ref{tab:user-oriented} and the model-oriented evaluation is reported in Table \ref{tab:model-oriented}. We note that the random baseline performs very bad on both evaluations, which shows the difficulty of the task and that randomly choosing aspects as explanations can barely reveal the reasons of the recommendations.

For the user-oriented evaluation, the results show that when applying the mask for fair comparison, CountER outperforms all the baselines on all the datasets in terms of $F_1$ scores. Moreover, CountER performs better than the baselines on precision in $90\%$ cases and on recall in $80\%$ cases. We note that our model has a very huge improvement than the baselines on Yelp dataset even without mask, and the reason may be that the Yelp dataset is much denser and has more reviews than other datasets so that the user's review bias has smaller impact. This also indicates that CountER has more advantages when the size of dataset increases.

For model-oriented evaluation, the mask limits CountER's ability to explain the model's behavior. However, no matter with or without the mask, our model has better performance than all the baselines according to $F_{NS}$ score. With mask, CountER has $15.63\%$ average improvement than the best performance of the baselines on $F_{NS}$. Without the mask, the average improvement is $38.49\%$.

Another observation is that the baselines commonly have higher PS scores, despite much lower on the PN scores. One possible reason is due to the mechanism of the matching based explanation methods. For an item on the recommendation list, they try to find well-aligned aspects where the user and the item both perform well. When computing the PS score, the baselines only reserve these well-aligned aspects in the aspect space for all the items, thus the recommended item will highly possibly still stay in the recommendation list,
which results in a high PS score. However, when computing the PN score, though the baselines remove all these aspects, the recommended item may still perform better on other aspects compared with other items and thus still remain in the recommendation list, which results in a lower PN score. This also sheds light on why the matching based methods may not discover the true reasons behind the recommendations.

\begin{figure*}[t]
    \centering
    \includegraphics[width=1\linewidth]{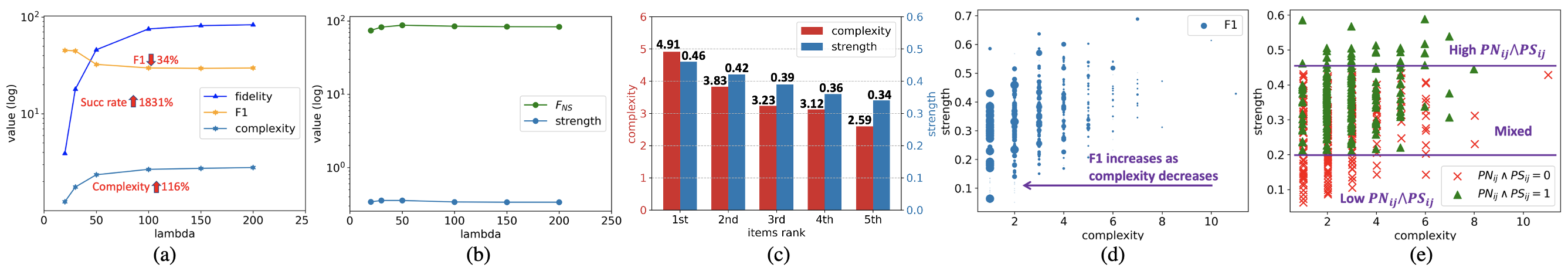}
    \vspace{-20pt}
    \caption{Ablation Studies. (a) Change of $F_1$ score, explanation complexity and fidelity w.r.t. $\lambda$. (b) Change of explanation strength and $F_{NS}$ score w.r.t. $\lambda$. (c) Average explanation complexity/strength for items at different positions. (d) Relationship between $F_1$ score with explanation complexity/strength. (e) Distribution of $PN_{ij}$ and $PS_{ij}$ with explanation complexity/strength.}
    \label{fig:ablation}
    \vspace{-10pt}
\end{figure*}

\vspace{-10pt}
\subsection{Influence of Relaxation Weight $\lambda$}
\label{sec:lambda}
Since $\lambda$ is applied on the hinge loss $L(s_{i, j_\Delta}, s_{i, j_{K+1}})$ (Eq.\eqref{eqn:optimizable} and \eqref{eqn:single_feature}), a larger $\lambda$ emphasizes more on the explanation strength and reduces the importance of complexity. 
As shown in Figure \ref{fig:ablation}(a), when $\lambda$ increases, the model successfully generates explanations for more items, but the explanation complexity also increases because more aspects are needed to explain those ``hard'' items. However, the higher $\lambda$ is, the worse our model performs on the user-oriented evaluation. Besides, Figure \ref{fig:ablation}(b) shows the explanation strength does not change with $\lambda$. This is because the $\alpha$ in the hinge loss controls the required margin on the ranking score to flip a recommendation. Additionally, we notice that the model-oriented performance is also independent to $\lambda$, which is the same as the explanation strength. 

Based on the above results, we hypothesize that the user-oriented performance may be related to the explanation complexity, while the model-oriented performance may be related to the explanation strength. This hypothesis is justified in Section \ref{sec:EC and ES}.

\subsection{Explanation Complexity and Strength}
\label{sec:EC and ES}
In this section, we study the Explanation Complexity and Explanation Strength. In Section \ref{sec:discussion on complexity}, we discussed that the difficulty of removing the items in the top-$K$ list are different based on their positions. Figure \ref{fig:ablation}(c) shows the mean complexity and strength of the explanations for items at different positions (i.e., from the first to the fifth). Generally speaking, it requires $1.59$ more aspects
to remove the items at the first position from the recommendation list than the items at the fifth position. Because they require larger change to be removed. This is in line with the fact that the strongly recommended items have more reasons to be recommended.

In Figure \ref{fig:ablation}(d), we illustrate the relationship between $F_1$ score and explanation complexity/strength. The scale of the marker points represent how large the $F_1$ score is. It shows that the user-oriented evaluation score $F_1$ decreases as the complexity increases, meanwhile, $F_1$ is relatively independent from the explanation strength. 

On the contrary, in Figure \ref{fig:ablation}(e), we plot the distribution of the explanations that are both necessary and sufficient (i.e., $\text{PN}_{ij} \wedge \text{PS}_{ij} = 1$), as well as the distribution of explanations that are either unnecessary or insufficient (i.e., $\text{PN}_{ij} \wedge \text{PS}_{ij} = 0$). We can see that as the explanation strength increases, more and more portion of the explanations are both necessary and sufficient. However, this tends to be irrelevant with the complexity of the explanations.

These observations are important because: 1) They indicate that the Explanation Complexity and Explanation Strength are two orthogonal concepts. Both of them are very important since the complexity is related to the coverage on the user's preference and the strength is related to the model's mechanism. 2) It legitimatizes the Occam's Razor Principle and further justifies the motivation of our CountER model, which is to extract both simple (low complexity) and effective (high strength) explanations for recommendations.





\section{Conclusions and Future Work}

In this paper, we proposed CountER, a counterfactual explainable recommendation framework, which generates explanations based on counterfactual changes on item aspects. Counterfactual reasoning is still in early stages for explainable recommendation, which has a lot of room for further explorations. For instance, CountER only explored changes on the item aspects. However, we can also explore counterfactual changes on other various forms of information such as images and textual descriptions.
The essential idea of CountER is also suitable for explainable decision making over knowledge graphs or graph neural networks, which are very promising directions to explore in the future.
\label{sec: conclusions}

\section*{Acknowledgments}
We appreciate the valuable feedback and suggestions of the reviewers. This work was supported in part by NSF IIS-1910154 and IIS-2007907. Any opinions, findings, conclusions or recommendations expressed in this material are those of the authors and do not necessarily reflect those of the sponsors.

\bibliographystyle{ACM-Reference-Format}
\bibliography{main_bib}

\end{document}